\documentclass[10pt,journal]{IEEEtran}

\IEEEoverridecommandlockouts

\usepackage{tikz}
\usepackage{amsmath}
\usepackage{lipsum}

\usepackage{filecontents}
\usepackage{balance}
\usepackage{adjustbox}
\usepackage{algorithm}
\usepackage{amsmath}
\usepackage[noend]{algpseudocode}
\usepackage{xcolor}
\usepackage{url}

\usepackage{subfigure}
\usepackage{balance} 
\usepackage{caption}

\usepackage{colortbl}

\usepackage[utf8]{inputenc}
\usepackage[english]{babel}

\def\BibTeX{{\rm B\kern-.05em{\sc i\kern-.025em b}\kern-.08em
		T\kern-.1667em\lower.7ex\hbox{E}\kern-.125emX}}

\newcommand{\ie}{{\em i.e., }}
\newcommand{\eg}{{\em e.g., }}

\newcommand{\myverb}{\fontsize{9}{48}\usefont{OT1}{lmtt}{b}{n}\noindent }

\ifCLASSINFOpdf
\else
\fi

\begin{document}

\title{A Survey on Enterprise Network Security: \\Asset Behavioral Monitoring and \\Distributed Attack Detection}

\author{
	Minzhao~Lyu,
	Hassan~Habibi~Gharakheili,	
	and~Vijay~Sivaraman
	
	\IEEEcompsocitemizethanks{
		\IEEEcompsocthanksitem  M. Lyu, H. Habibi Gharakheili and V. Sivaraman are with the School of Electrical Engineering and Telecommunications, University of New South Wales, Sydney, NSW 2052, Australia (e-mails: minzhao.lyu@unsw.edu.au, h.habibi@unsw.edu.au, vijay@unsw.edu.au). 
		
		\protect
		
	}
}

\IEEEtitleabstractindextext{%
\begin{abstract}
Enterprise networks that host valuable assets and services are popular and frequent targets of distributed network attacks. In order to cope with the ever-increasing threats, industrial and research communities develop systems and methods to monitor the behaviors of their assets and protect them from critical attacks.
In this paper, we systematically survey related research articles and industrial systems to highlight the current status of this arms race in enterprise network security.
First, we discuss the taxonomy of distributed network attacks on enterprise assets, including distributed denial-of-service (DDoS) and reconnaissance attacks.
Second, we review existing methods in monitoring and classifying network behavior of enterprise hosts to verify their benign activities and isolate potential anomalies.
Third, state-of-the-art detection methods for distributed network attacks sourced from external attackers are elaborated, highlighting their merits and bottlenecks.
Fourth, as programmable networks and machine learning (ML) techniques are increasingly becoming adopted by the community, their current applications in network security are discussed.
Finally, we highlight several research gaps on enterprise network security to inspire future research.
\end{abstract}
}

\maketitle
\IEEEdisplaynontitleabstractindextext
\IEEEpeerreviewmaketitle

\section{Introduction}\label{sec:Intro}
Enterprises such as universities and research institutes host critical data and offer publicly accessible services through their networks. 
Thus, they often become popular targets of distributed network attacks that actively probe asset vulnerabilities and paralyze their services.
With practical defense appliances (\eg firewalls and intrusion detection systems) employed by IT departments of enterprises, network attacks are becoming well distributed in sources and agile in attacking patterns to bypass such static detection and increase their effectiveness.
To be more specific, a sophisticated network attack usually employs hundreds and thousands of botnet devices spread across geolocations and diversified in types (\eg Internet-of-Things, laptops, and compromised servers); each may send malicious traffic with changing patterns and protocols.
Some popular and large-scale DDoS attacks \cite{CloudFlareFamousDDoS} include, but are not limited to: Amazon AWS became the target of a massive Terabits-level DDoS attack sourced from hijacked CLDAP servers in 2020; Github suffered from a Memcached protocol-based DDoS attack in 2018; during Rio 2016 Summer Olympics, critical servers of official Olympics organizations as well as Brazilian banks and telcos \cite{HSolomon2020}) were targeted by sustained distributed network attacks with mixed traffic types such as TCP-SYN, UDP reflection, DNS, CHARGEN (character generator protocol), NTP, and SSDP sourced from millions of compromised devices (\eg IoTs) across the globe \cite{OlynmicAttack2018}. 
Successful distributed network attacks lead to service failures, disruptions, and reputation degradation.

\begin{figure*}[!t]
	\begin{center}
		\includegraphics[width=1\textwidth]{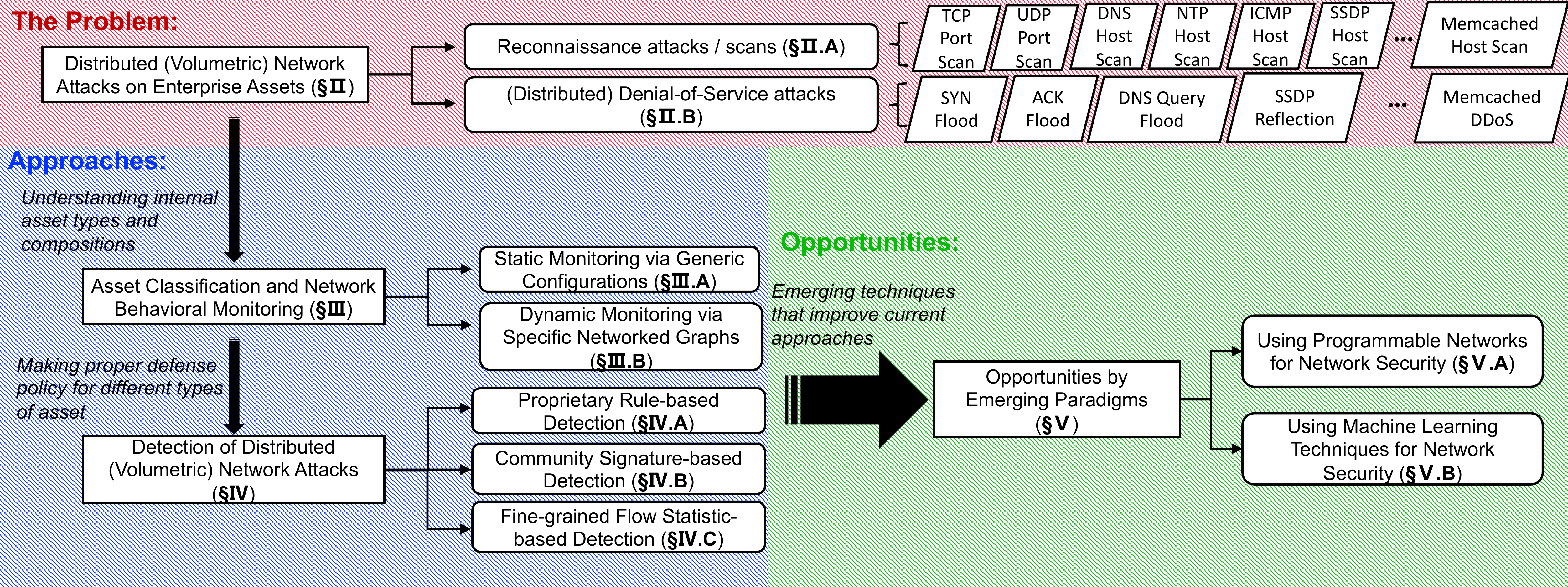}
		\caption{Key topics covered in this survey.}
		\label{fig:keyLogistic}
	\end{center}
\end{figure*}

Distributed attacks on enterprise networks often consist of two phases, namely reconnaissance attacks (also known as scans) to discover the vulnerability of networked assets and distributed denial-of-service (DDoS) attacks that paralyze the targeted victims that are discovered by malicious actors.
To cope with the threats, enterprise IT departments are expected to track the devices within their networks to ensure their expected behaviors and enforce attack defense mechanisms that can effectively detect and mitigate attacks on their networked assets without impacting legitimate communications.

There are many mature products for monitoring the network behaviors of enterprise assets and providing protections against distributed attacks via static configurations, such as next-generation-firewall (NGFW) appliances and intrusion detection systems (IDS). 
These static solutions are practical to be used in high-throughput enterprise networks. Still, they are ineffective in providing precise results (\eg differentiating distributed attackers and malicious flows from their benign counterparts). Therefore, it is not surprising that the consequential attack mitigation measures (\eg randomly dropping packets to the victim) introduce non-negligible collateral damages on benign traffic \cite{CDietzelCoNEXT2018}. 
For instance, typical next-generation-firewalls (NGFW) require users to configure rules that specify the list of focused enterprise assets and the corresponding defense strategies. Such methods effectively protect certain critical assets by tracking their network activities of several traffic types but fail to capture unknown and complex threats from hosts operated by sub-departments, staff, and visitors. Moreover, the static nature of such methods limits their capabilities in detecting emerging attacks with dynamic and stealthy traffic patterns \cite{MZhangNDSS2020}. 

Legacy static solutions introduce blind spots likely exploited by malicious actors and agile attackers. Research communities have developed dynamic telemetry methods for network monitoring via flow-level statistics and networked graph structures to address this problem.
Those methods can provide fine-grained statistics to track each network flow between enterprise assets and external hosts without leaving any blind spot. However, maintaining fine-grained flow-level telemetry unavoidably introduces high computational overheads. Therefore, they are not scalable for large enterprise networks with hundreds and thousands of hosts that exchange millions of concurrent flows.

Recent developments in two emerging paradigms, namely Programmable Networks and Machine Learning (ML), offer promise to improve the flexibility of network monitoring and accuracy of attack detection. 
Generally speaking, programmable networking covers two main areas: network function virtualization (NFV) and software-defined networking (SDN). It changes the static nature of network traffic processing often carried out by proprietary legacy hardware and middleboxes. Instead, dynamic network functions on generic servers and programmable switches are used to achieve high responsiveness and real-time orchestrations. Researchers have leveraged this technology to overcome the challenges of legacy network monitoring and protection in various use cases, such as real-time defense orchestration for ISP network \cite{SKFayazSec2015} and elastic control of virtual firewalls \cite{JDengNDSS2017}. These inspire the development of solutions to the current problems of enterprise network security.
On the other hand, recent advances in ML techniques that help obtain data-driven models to make accurate predictions on statistical attributes have proven their supremacy in many disciplines, such as computer vision and language recognition. Despite some of the practical challenges in applying ML methods to network security \cite{RSommerSP2010}, researchers have successfully employed ML algorithms to make reliable security inferences from various types of network telemetry (\eg system logs or packet headers) in a variety of scenarios (\eg IoT attack or SSH brute-forcing). We believe that their trials and efforts provide us with valuable lessons to address issues in asset classification and attack detection accurately and precisely.

This survey systematically reviews related research articles and industry practices, providing comprehensive insight into current developments, challenges, and future directions of asset management and distributed attack detection in enterprise network security.
Unlike prior surveys that broadly studied certain attack types and defense mechanisms, we focus on a narrow aspect of distributed volumetric network attacks and their countermeasures applicable to enterprise networks. In addition, we review the potential and challenges of improving the state-of-the-art in two emerging paradigms (\ie programmable networks and ML). To this end, we summarize the main topics covered by this survey as follows, which are also visually shown in Fig.~\ref{fig:keyLogistic}.  \textbf{First}, in \S\ref{sec:SurveyDistributedAttack}, we highlight the diversity and variety of distributed network attacks including reconnaissance scans and distributed denial-of-service (DDoS) attacks; \textbf{second}, in \S\ref{sec:SurveyNetworkAssetTracking}, we discuss the current development of enterprise networked asset classification and behavioral monitoring via either static or dynamic methods; \textbf{third}, in \S\ref{sec:SurveyDistributedAttackDefense}, enterprise distributed attack detection systems using proprietary rules, community signatures, and fine-grained flow statistics are surveyed; \textbf{fourth}, in \S\ref{sec:SurveyEmergingTechniques}, opportunities introduced by the two emerging paradigms, \ie flexibility by programmable networks and accuracy by machine learning are discussed as to inspire future researches.
Relevant surveys (but on other aspects of network security) are discussed in \S\ref{sec:RelatedWorks}. We highlight several research gaps as valuable future directions in \S\ref{sec:discussion}, and conclude this survey in \S\ref{sec:conclusion}.

\begin{figure*}[!t]
	\begin{center}
		\includegraphics[width=0.9\textwidth]{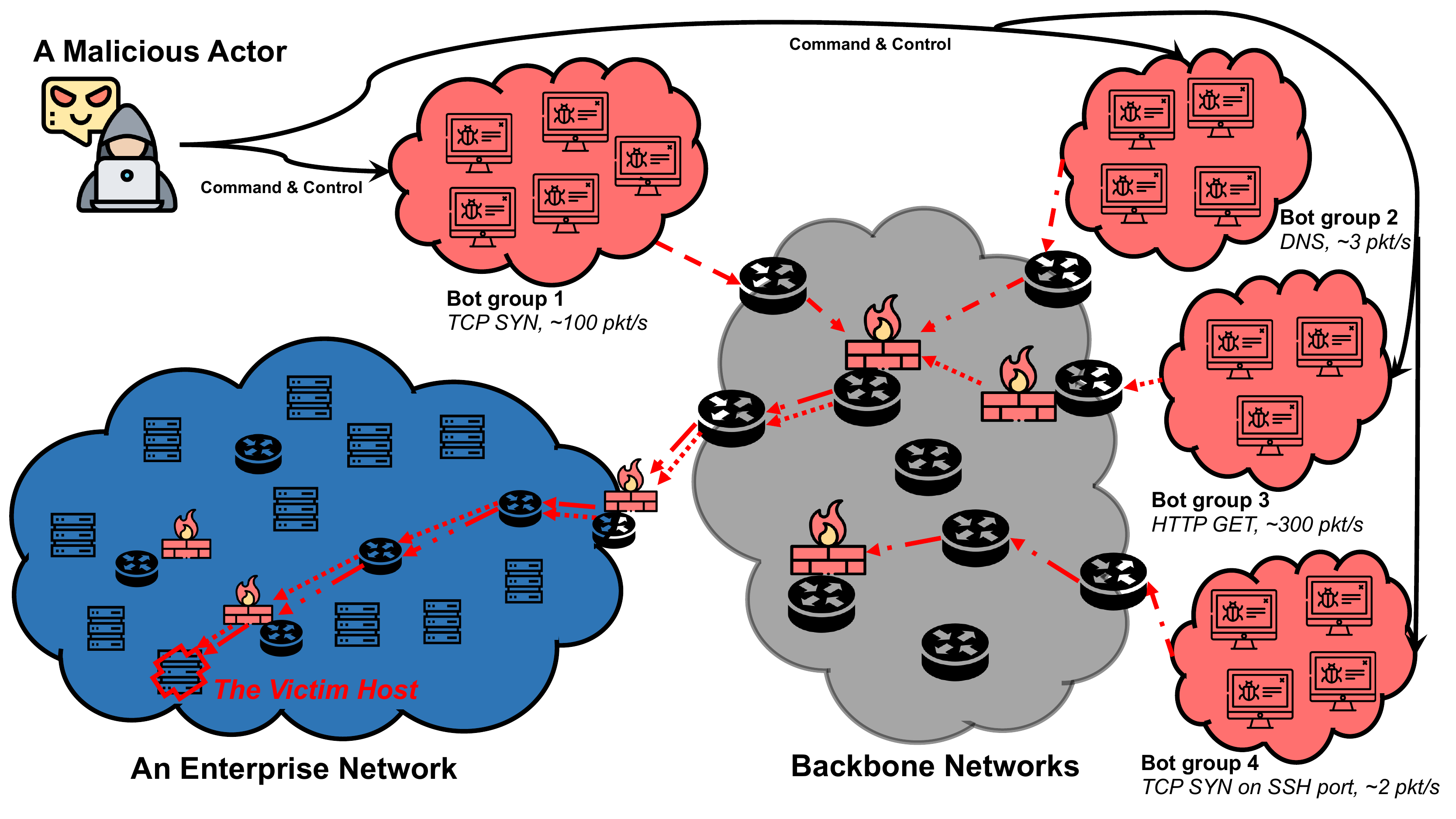}
		\caption{A visual example of distributed network attacks on a victim inside an enterprise.}
		\label{fig:anatomyOfAttack}
		\vspace{-5mm}
	\end{center}
\end{figure*}

\section{Distributed Network Attacks \\on Enterprise Assets}\label{sec:SurveyDistributedAttack}

Network attacks that probe, congest, or paralyze enterprise assets such as public-facing servers are becoming distributed in sources, versatile in traffic patterns, and diverse in underlying mechanisms \cite{JMirkovicCCR2004,ALakhinaCCR2004,NAgrawalCST2019}.
Such attacks often occur sequentially -- an enterprise asset is first examined for its availability and known vulnerabilities through a reconnaissance attack (\ie host or port/service scans), followed by (distributed) denial-of-service (DoS or DDoS) attacks. 

Large-scale scans and denial-of-service are often conducted in a distributed manner from a single source to (a) increase their effectiveness and (b) make it difficult for defense systems to detect and/or mitigate. 
Distribution is typically achieved by recruiting botnets, consisting of massive compromised devices like personal computers, powerful workstations, public-facing servers, or compromised IoT appliances \cite{FFreilingESORICS2005,GGuBotSniffer2008,FTegelerCoNEXT2012,MThomasWWW2014,YNadjiCCS2013}.
To avoid detection, malicious actors often split an attack into small segments, each performed by a single bot device.
For example, in a powerful but stealthy DDoS attack, each bot device only generates low-rate traffic across a variety of protocols \cite{THeinrichPAM2021}, making it difficult to be distinguished from benign instances. 
It is practically challenging to precisely identify all attack sources \cite{MAbuRajabIMC2006} and block them.
We show a visual example of distributed network attacks in Fig.~\ref{fig:anatomyOfAttack}, where a malicious actor commands and controls four distributed bot groups to attack a victim residing within an enterprise network. Each group uses a different traffic type and rate so that a reasonable fraction of the entire attack traffic (sent by bot groups 1 and 2 in Fig.~\ref{fig:anatomyOfAttack}) can successfully bypass defense appliances. Note that most commercial firewalls operational in backbone and enterprise networks are not optimally tuned to detect stealthy malicious traffic on the path before it hits the victim.

\subsection{Reconnaissance Attacks}

Malicious actors use reconnaissance attacks (also known as scans) to construct their knowledge of targeted hosts and services (ports). Those attacks probe the availability of enterprise-connected hosts and discover their potential vulnerabilities 
\cite{ATundisARES2018}. 
The discovered hosts may not only become victims but may also be exploited as attack amplifiers/reflectors to paralyze other victims.
For example, a discovered NTP server with high reflection capability (\ie generate response packets with a size larger than that of the received requests) can be used to amplify attack volume in reflection-based DDoS attacks \cite{MKuhrerSec2014,MLyuWISEC2017}.

Apart from malicious purposes, security researchers also develop tools to identify potential cyber threats enterprises face, such as open ports and vulnerable services that could be exploited in network attacks. 
For example, \textit{Nmap} \cite{AOrebaugh2011} is developed as a comprehensive scanning tool to discover active hosts and ports (\ie services). 
To increase the speed and effectiveness of scans, the authors of \textit{Zmap} \cite{ZDurumericSec2013} optimized the scanning process by tuning the probing rate, pre-connection state, and re-transmission, which can probe the entire IPv4 space within 45 minutes.
Scanning techniques have evolved to become scalable at 10 Gbps throughput \cite{DAdrianWOOT2014} and can perform vulnerability scans towards protocol banners through user queries \cite{ZDurumericCCS2015}.
Reconnaissance attacks have also been studied for certain scan types, such as critical cyber-physical infrastructures \cite{FYarochkinPRISDC2013} and DNS utilities \cite{QHuDSN2018}.

To combat reconnaissance attacks, researchers have developed methods, such as tracking port scanners on the IP backbone \cite{ASridharanLSAD2007}, detecting subtle port scanning via interactive visualization \cite{WWangRIIT2014}, disrupting reconnaissance attacks via address mutation \cite{JJafarianTIFS2015}, constructing distributed network telescope to capture scanners \cite{PRichterIMC2019}, and optimizing backscatter \cite{KFukudaTNET2017} technique for scan detection in massive IPv6 address space \cite{KFukudaIMC2018}.
However, according to \cite{HHeoASIACCS2018,ZDurumericSec2014}, only a few enterprises have practically adopted robust defensive measures. Thus, service and host scans are still prevalent on the Internet, exposing service and device vulnerabilities (\eg Linksys routers, OpenSSL, and NTP). Consequently, exposed hosts may be exploited by malicious actors on the Internet to generate/reflect attacks or become direct victims in the future.

\subsection{Distributed Denial-of-Service Attacks}

As already shown in Fig.~\ref{fig:anatomyOfAttack}, malicious actors may choose to flood their target victim directly from botnet devices using various techniques or protocols \cite{THeinrichPAM2021} (\eg HTTP, ICMP, and TCP-SYN).
Also, they may choose to launch a reflection-based DDoS with larger attack volumes. For example, bot devices send packets with the spoofed source IP address (of the ultimate victim) to the discovered reflectors (\eg DNS and NTP servers); these reflectors will then respond to the victim with larger packet sizes.

DDoS attacks are becoming more complex, distributed, and agile. The existing research literature extensively studied the characteristics of various DDoS attacks.
First, according to \cite{GGuSecurity2008}, DDoS attacks are becoming complex in protocols and traffic types. The participant botnets are likely to be independent. Such patterns make it challenging for defenders to isolate malicious traffic and attack sources.
Second, the increasing adoption of cyber-physical devices (\eg IoTs) brings new vulnerabilities and expands attack surfaces yet to be addressed \cite{FLoiIoTSP2017}. Therefore, a growing number of IoT devices connected to the Internet are compromised as a botnet, enabling more powerful and frequent DDoS attacks on a global scale \cite{AWelzelEuroSec2014}.
For example, in late 2016, Mirai \cite{MAntonakakisSec2017}, an IoT malware that hijacked hundreds of thousands of IoT devices, has led to unprecedented DDoS attacks globally. During an attack, each compromised IoT device generates malicious traffic at a low rate, making them hard to be differentiated from benign traffic.
Third, DDoS attacks are becoming more dynamic and agile in their activity patterns to evade detection. As pointed out in \cite{AWangTDSC2020}, they are usually launched with changing temporal and spatial patterns to bypass detection, which makes them quite effective against static rule-based and signature-based detection methods. Botnets of different families also work collaboratively. A given bot might adapt its attacking strategy provided by different malware families \cite{WChangASIACCS2015}.
Finally, the concept of DDoS-as-a-service is becoming popular as it lowers the barrier to generating a distributed attack effectively \cite{MKaramiLEET2013}. Botnet owners can lease their controlled devices for financial benefits, so malicious actors with fewer resources (\eg controlled bot) can rent their large botnet to launch powerful attacks.

\subsection{Highlights of Distributed Network Attacks}

We now summarize three key highlights in this section.

First, network attacks such as DoS and scans are becoming: (a) ``\textit{distributed}'' by recruiting botnets to generate attack traffic from different logical sources (\eg ASes, subnets) and physical geolocations, (b) ``\textit{complex}'' by leveraging a wide range of protocols and vulnerabilities, and (c) ``\textit{dynamic}'' by shifting active bot groups or traffic patterns randomly. All the above characteristics increase the difficulties in effectively detecting distributed attacks.

Second, potential vulnerabilities of network-connected hosts (\eg BYOT devices, enterprise servers, or IoTs) may be identified and exploited by malicious scripts (\eg URLs contained in phishing emails) or malware. Such compromised devices are used as bots to perform further infections within their local network or participate in attacks on other networked assets. Therefore, continuously monitoring network traffic behaviors and enforcing appropriate security management are essential for network operators (IT and cyber departments). We will discuss in  \S\ref{sec:SurveyNetworkAssetTracking} some of the tools and techniques for asset network behavioral monitoring.

Third, apart from malware infections and misuse, assets such as servers and databases within an enterprise network may be direct targets of distributed attacks. During such attacks, public-facing servers may not be able to respond to benign requests of external clients if their networking and computational resources get exhausted. In addition, network vulnerabilities of internal non-critical enterprise hosts may become exposed to external hackers for further cyber-crimes. Therefore, defending against distributed attacks on enterprise assets is critical for security operations. In \S\ref{sec:SurveyDistributedAttackDefense}, we will elaborate on state-of-the-arts enterprise attack detection systems and mechanisms.

\section{Asset Classification and \\Network Behavioral Monitoring}\label{sec:SurveyNetworkAssetTracking}

Obtaining real-time visibility into assets and their behaviors is essential to combat the increasing number of distributed network attacks targeting or utilizing enterprise assets. IT\footnote{information technology}, OT\footnote{operational technology}, and cybersecurity teams need tools to classify connected assets based on their role (\eg web server, DSN server, camera, personal computer), ensuring asset activities conform to their role's patterns (profile). With asset profiles clearly modeled, appropriate security policies (\eg segmentation, access rules) can be applied to the network, and certain attacks can be prevented or at least detected easier.

\begin{table}[t!]
	\caption{Classifying host types in a large enterprise network by DNS names \cite{MLyuCN2022}.}
	\centering
	\renewcommand{\arraystretch}{1.2}
	\begin{adjustbox}{max width=1\textwidth}
		\begin{tabular}{|l|c|l|c|c|c|c|c|c|}
			\hline
			\rowcolor[rgb]{ .906,  .902,  .902}	\textbf{Asset type}             & \textbf{\# hosts} \\ \hline
			Website server 		  & 		61\\ \hline
			Authoritative name server   & 15\\ \hline
			VPN gateway server   & 13 	\\ \hline
			Remote computing platform   & 16	\\ \hline
			File storage server    & 14 	\\ \hline
			Mail exchange server    & 18   \\ \hline
			DNS recursive resolver    & 7 \\ \hline
			Web proxy   & 4	\\ \hline
			NAT gateway    & 256\\ \hline
			Personal computer and BYOTs  & 1,961 \\ \hline
			Other unclassified (minor) types & 18,920 \\ \hline
		\end{tabular}
	\end{adjustbox}
	\label{tab:HostFunctionality}
\end{table}

However, profiling asset behaviors is a nontrivial task as enterprise hosts come with diverse and complex functionalities and behaviors. For example, an enterprise can have servers of various types that serve internal or external clients; visitors and staff may have their personal devices (\eg mobile phones and laptops) connected through wireless gateways, and IoTs such as smart cameras and sensors may also be installed in a typical enterprise network \cite{ASivanathanTMC2019}. 
Let us take a look at Table~\ref{tab:HostFunctionality}, which lists popular types of networked hosts (top ten rows) identified by their enterprise DNS names in a large university network, studied by work in \cite{MLyuCN2022}. As shown by the last row of Table~\ref{tab:HostFunctionality}, there are many other unclassified and less-popular host types, such as LDAP server and Redis proxy, which are often hard to enumerate. We note that those identifiable assets (by their domain name) are only accountable for less than 10\% of active hosts in the enterprise, and the functionality of other 90\% hosts is mostly unknown to the university network managers and administrators.

Connecting many heterogeneous devices will inevitably introduce challenges to network management, operation, and security teams.  Devices owned and managed by visitors and/or staff may come infected by malware and hence start conducting malicious activities \cite{MEslahiISCAIE2014} upon arrival to the network, which may go undetected by security tools and appliances \cite{ROgieCEM2016,HAlmohriTDSC2016}. Also,  inaccurate access policies and configurations (\eg public-facing servers) may give external attackers opportunities to compromise less secure internal hosts for malicious purposes. Organizations like universities and research institutions often have relatively unfettered networks, allowing subdivisions and departments to configure their own IT infrastructures. This makes the problem even more pronounced as asset visibility gets relatively poor.

Many solutions have been developed by industry (\eg \cite{Manageengine2022,CiscoFirewallBestPractices2018,Dynatrace2022}) and academia (\eg \cite{MLyuCN2022,TKaragiannisSIGCOMM2005,EGlatzVizSec2010,JJPfeifferWWW2014}) to classify roles and/or monitor network behaviors of individual assets. Existing methods can be categorized into either static configurations/databases that record high-level characteristics (\eg role/class/model) of the connected devices or dynamic graphs obtained from passive traffic monitoring that capture communication patterns individual networked hosts display.

\subsection{Static Monitoring via Generic Configurations}

\begin{figure*}[t!]
	\begin{center}
		\mbox{
			
			\subfigure[A website server.]{
				{\includegraphics[width=0.79\textwidth]{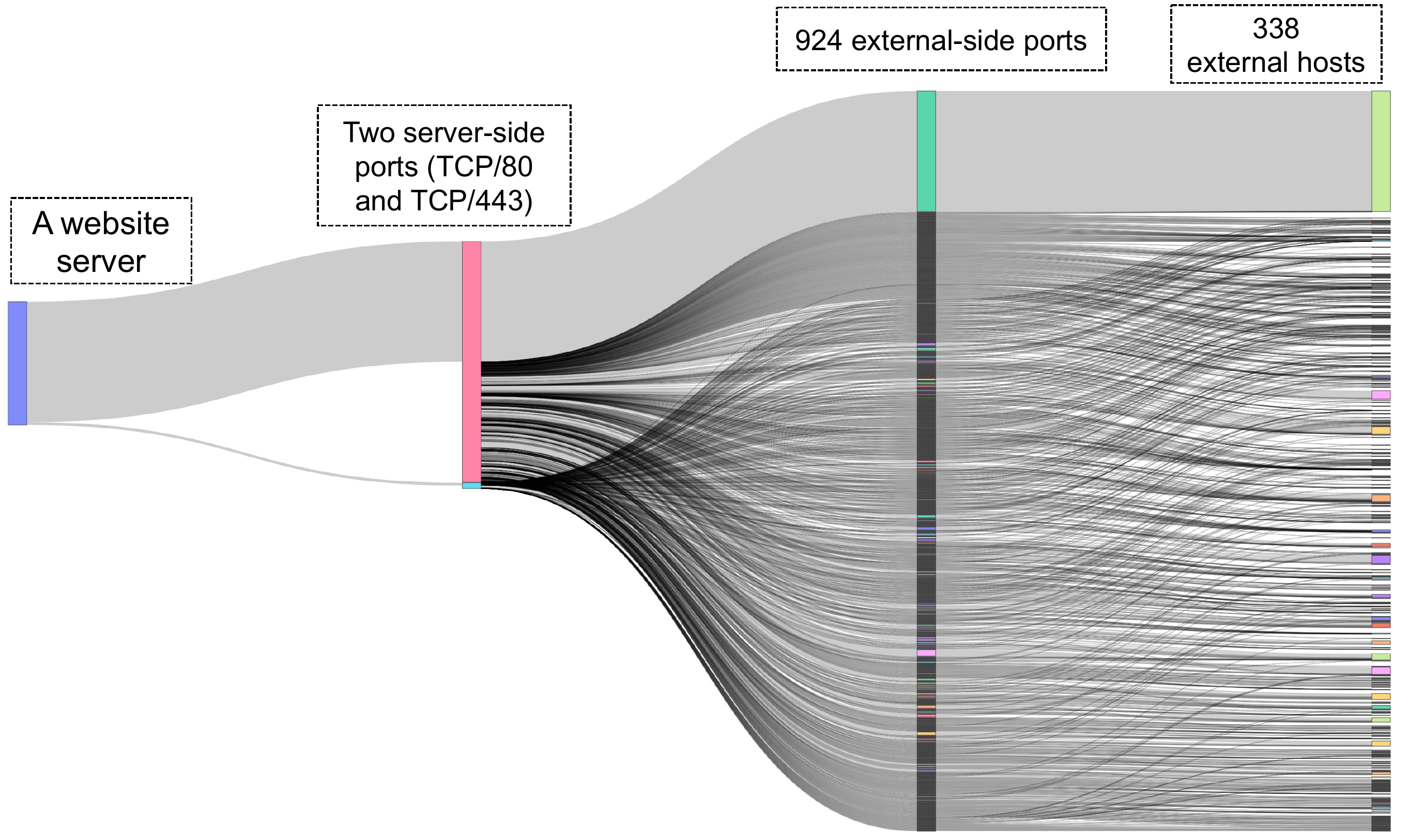}}\quad
				\label{fig:webserver}
			}
		}
		\mbox{
			
			\subfigure[A DNS recursive resolver.]{
				{\includegraphics[width=0.79\textwidth]{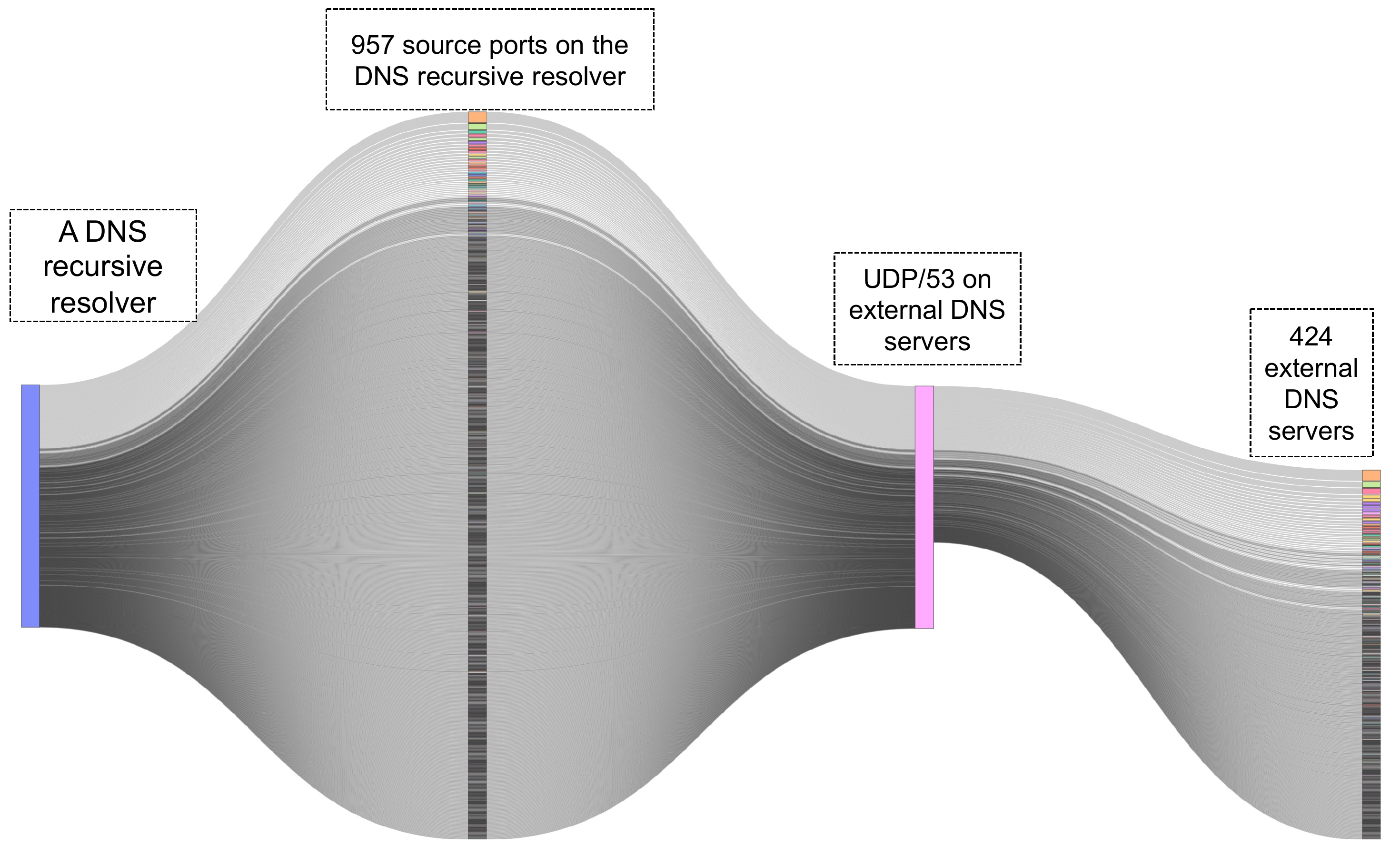}}\quad
				\label{fig:dnsresolver}
			}	
		}	
		\vspace{-2mm}
		\caption{Sankey diagrams illustrating network behavioral profiles of two representative enterprise assets: (a) a website server, and (b) a DNS recursive resolver, using 1000 flows of each networked asset for visualization purpose.}
		\label{fig:sankeyplot}
		\vspace{-6mm}
	\end{center}
\end{figure*}

Current practical solutions for the management and classification of enterprise networked assets primarily rely on static and relatively generic characteristics (\eg tables containing a list of device hostnames, their VLAN, Operating System, IP address, or perhaps their role) without capturing their behavioral characteristics like what is shown in Fig.~\ref{fig:sankeyplot}. For example, firewall appliances are often configured by access control lists (ACL) and rules that keep static information of internal hosts such as VLAN ids, device categories, and/or user groups \cite{PaloAltoFirewall30202018}; DHCP and DNS servers maintain system logs containing device names and their IP addresses \cite{DNSRecursiveResolverSecurity2017}; and, other specialized commercial platforms managing enterprise assets are configured by lists supplied (often manually) by the IT department \cite{Gartner2022}.

Ideally, an IT department equipped with full knowledge of assets connected to their network will be able to segment the network and enforce strict access configurations to prevent unintended communications to/from networked devices on the network \cite{CiscoFirewallBestPractices2018}. Network traffic not conforming to those configurations will be marked as abnormal, thus, triggering further defense actions such as alerts and mitigation. For example, according to best practices of the Microsoft firewall \cite{Microsoft2022}, inbound port or service rules could be enforced so that the border firewall blocks all non-HTTPS traffic towards an enterprise HTTPS server or inbound DNS packets are only allowed if their destinations are enterprise DNS servers.
To protect a critical asset operational within an enterprise (say, a corporate website server), the network administrator may choose to set up an upper bound rate limit via its traffic sharper platform for that asset. Whenever the asset receives traffic rates higher than the allowed limit, the management system (\ie shaper and/or firewall) will partially or fully drop those inbound packets, preventing a potential volumetric attack on that specific host.

Static configurations enable administrators to manage and monitor their enterprise assets by specifying relatively high-level network profiles. However, with the explosive growth of network applications communicating via a variety of protocols in conjunction with the adoption of IoT/OT devices with heterogeneous behaviors, populating and maintaining generic, high-level configurations and policies become increasingly difficult for enterprise IT departments, especially for those with loosely-federated networks \cite{ENgoupeIM2015}. 
As highlighted in \cite{PWangDCSW2009,AStarschenkoICUMT2015,AVoronkovCS2017,JGarciaAlfaroCS2013}, specifying policies for a large enterprise network with complex host composition is error-prone, and potential misconfigurations can impose high operational costs, such as fixing bugs and resolving conflicts.
This problem becomes even worse with the adoption of many and diverse BYOTs and IoTs \cite{HIsmailGCIoT2018}. Therefore, managing assets by simple (generic) configurations tailored for each device type inevitably leaves many blind spots and becomes impractical in most operational settings \cite{AWongJSAC2005,HBallaniSIGCOMM2007}.

Moreover, the behavior of networked hosts in modern enterprises can change in time \cite{MLyuCN2022}. For example, certain divisions or departments (engaged in projects with external stakeholders) may operate multiple services (\eg DNS and website) on a single machine and expose them to the public Internet, each with a distinct behavioral pattern -- some services may get terminated, and/or new services or functionalities may be added on-demand. As a result, static methods fall short of expectations \cite{AWedgburyICSCSR2015,FZhuPC2005,TBakhshiITA2015}. 

Motivated by some of challenges highlighted above, researchers have developed dynamic methods using specific networked graphs, which will be discussed next.

\subsection{Dynamic Monitoring via Specific Networked Graphs}

To obtain fine-grained visibility into  activities of connected hosts, prior works (will soon be discussed in this section) used networked graphs to characterize (profile) the behavior of various host types.
To motivate our discussion, let us consider Fig.~\ref{fig:sankeyplot}, which visualizes the flow graphs\footnote{These are constructed from data and models presented in \cite{MLyuCN2022}.} (in the form of Sankey diagrams) of two enterprise hosts connected to a university network (\ie a website server and a DNS recursive resolver).
The website servers often expose only two TCP ports (\ie {\myverb{TCP/443}} and {\myverb{TCP/80}}) \cite{MLyuCN2022} to the public Internet allowing for communication sourced from a wide range of TCP ports by external hosts, while the DNS recursive resolver sends traffic from arbitrary UDP ports targeting only  {\myverb{UDP/53}} operational on external DNS servers.

\subsubsection{Per-Host Classification}

Work in \cite{TKaragiannisSIGCOMM2005} uses graph structures to model network activities of each connected host at the IP address and transport-layer port levels. The authors profiled various types of networked hosts (\eg HTTP servers, DDoS attackers, and P2P clients), each with a unique transport-layer behavioral pattern. For example, the graph pattern of an FTP server consists of a large number edges destined to IP addresses, initiated from a wide range of port numbers connecting to two popular port numbers, namely {\myverb{TCP/20}} and {\myverb{TCP/21}} on the server. The authors developed a method to classify an unknown host by checking the similarity between its behavioral graph with that of known types.
To effectively monitoring communication patterns of hosts, work \cite{EGlatzVizSec2010} developed a tool that visualizes communication graphs for network operators helping them classify networked entities manually.
To fill the gap in legacy graphs that only capture flow profiles, researchers have developed relatively advanced graph structures that can model network communications with more descriptive features, such as attributed graph models in \cite{JJPfeifferWWW2014}. This graph structure is generated to capture both network topological properties (\eg connections between nodes) and correlated attributes on each graph edge which hold both computational efficiency through sampling techniques used in graph generation and accuracy when classifying host roles in real-world networks.

\subsubsection{Clustering Hosts and Modelling Group Interactions} 

Modeling the behavior of individual hosts can be challenging, especially at scale. Therefore, one may choose to reduce the dimension by focusing on clusters/groups of hosts. In such clustered graphs, hosts are often grouped and represented by their common communications behaviors, such as contacting a similar range of external hosts or residing in the same subnet. However, balancing the level of aggregation and visibility into the actual network often requires extensive tuning and optimization.
In \cite{PBhattacharyaIS2012}, the authors optimized communication graphs for a large network to achieving the optimal consumption of computational resources while having sufficient information captured in the graph to describe the evolution of a network attack.
Authors of \cite{JJuskoSPCN2013} used the method named ``connection graph analysis'' to discover cooperating hosts in P2P networks, which start from a single known P2P node in a network to discover and group other associated hosts progressively. The developed method was demonstrated to have short processing times in grouping all P2P hosts in large networks by processing their NetFlow streams.
Besides, statistical methods such as clustering algorithms are quite powerful for grouping and differentiating hosts based on their behavioral profiles in large-scale networked graphs. As an example, in \cite{KXuTNET2014}, the authors applied clustering algorithms to effectively identify groups of hosts that inherently belong to different application types on bipartite graphs describing communications between hosts.
As in \cite{JAhmadCN2016}, the authors clustered hosts within the same enterprise network that have strong inter-IP connectivity (\ie connecting to a similar range of hosts) for enterprise IT departments so that they could track the behavior of each identified group instead of individual hosts for scalability in network management.

\subsubsection{Detecting Host Anomaly from Graphs} 
Using networked graphs that describe host interconnectivity can be particularly useful for cybersecurity applications such as to detect malicious hosts in a network or identify clusters of botnet devices launching distributed attacks \cite{NPoolsappasitTDSC2012,LWangDAS2008,MAlbaneseDSN2012,MGonzalezTPS2017,HAlmohriTDSC2016,DEswaranKDD2018,PKalmbachSIGCOMM2019,IHassaanTIFS2021}.
For some recent examples, the work in \cite{HAlmohriTDSC2016} considers a large enterprise network with dynamic compositions and communication patterns of hosts, and hence becomes difficult to manage and secure. 
Therefore, the authors developed a probabilistic graph model to measure the success rate of an attack on a given network topology so that IT departments could optimize their attack detection policies and fix vulnerable network configurations.
\textit{SpotLight} \cite{DEswaranKDD2018} achieved accurate and responsive detection of anomalies in high-density graphs for IP communications near real-time. The anomalies (\eg port scans and DoS) are identified by the sudden changes in subgraphs consisting of a subset of nodes and edges from the networked graph. The authors leveraged randomized sketching algorithms to make cost-effective inferences with optimal memory consumption.
Similarly, \textit{Noracle} \cite{PKalmbachSIGCOMM2019} detect anomalous behavioral changes of individual hosts in network graphs using stochastic block models, which could detect hosts with deviated behaviors (\eg connecting to unusual hosts) compared with other hosts in the same cluster.
Whereas \textit{TRACE} \cite{IHassaanTIFS2021} builds a distributed enterprise-wide communications graph tracking information from both network connectivity (\eg IP address and port number) and involved device system calls (\eg application name and process ID) between enterprise hosts for advanced persistent threat (APT) detection.

\subsection{Highlights}
In summary, configuring static policies on middleboxes like firewalls is the de facto method by the current industry practices for managing (selected) networked assets.  Such methods are practical computationally, as they often maintain lightweight data structures for specific groups of managed entities, those with critical values to and/or functions for an enterprise. This method prioritizes practical deployment but makes it difficult to gain fine-grained visibility (\eg, at the flow level) and effectively classify host behaviors that are often dynamic or unknown to IT departments. 

On the other hand, dynamic monitoring with specific networked graphs is proven to be effective in providing comprehensive visibility into network traffic so that IT departments can effectively classify connected assets and detect potential anomalies. However, using complex graphs incurs high computational costs that make such methods impractical for deployment in large enterprise networks with many diverse, active hosts and concurrent communication flows.

\section{Detecting Distributed Network Attacks on Enterprise Hosts}\label{sec:SurveyDistributedAttackDefense}
Detecting distributed attacks (\ie DDoS and reconnaissance) is critical for enterprise network operations. To date, the cybersecurity research community has developed solutions to detect various distributed network attacks.
For those attacks aiming to congest the Internet link of an enterprise network by sending Gbps or even Tbps malicious traffic to enterprise hosts, handling the attack at ISP levels (close to source and in-transit) appears to be the most effective option \cite{STZargarCST2013,JMirkovicNCA2003,JMirkovicTDSC2005,GOikonomouACSAC2006,SKFayazSec2015}. For distributed attacks targeting certain enterprise assets, which is the focus of this survey, detection mechanisms employed by the target enterprise (close to victim/destination) are proven to be more effective \cite{STZargarCST2013}. Therefore, enterprise IT departments usually set up inline security middleboxes near their network edges, sitting in between their internal private network and the public Internet. To this end, monitoring and/or detection policies can be developed and enforced for, say, each of the critical enterprise servers \cite{WMichaelTCS2010} that is attractive to potential attackers. Such detection solutions, typically employed by enterprise IT departments, can be categorized into three types: proprietary rules, community signatures, or (flow-level) statistical models, which are comprehensively reviewed in this section.

\subsection{Proprietary Rule-Based Detection}\label{sec:rule-based}

Rule-based distributed attack detection, which allows users to configure their security policies from a list of rules defined by the appliance manufacturer or developer, is widely used by the enterprise security industry.

\begin{figure*}[t!]
	\begin{center}
		\mbox{
			\hspace{-4mm}
			\subfigure[Reconnaissance attack.]{
				{\includegraphics[width=0.985\textwidth]{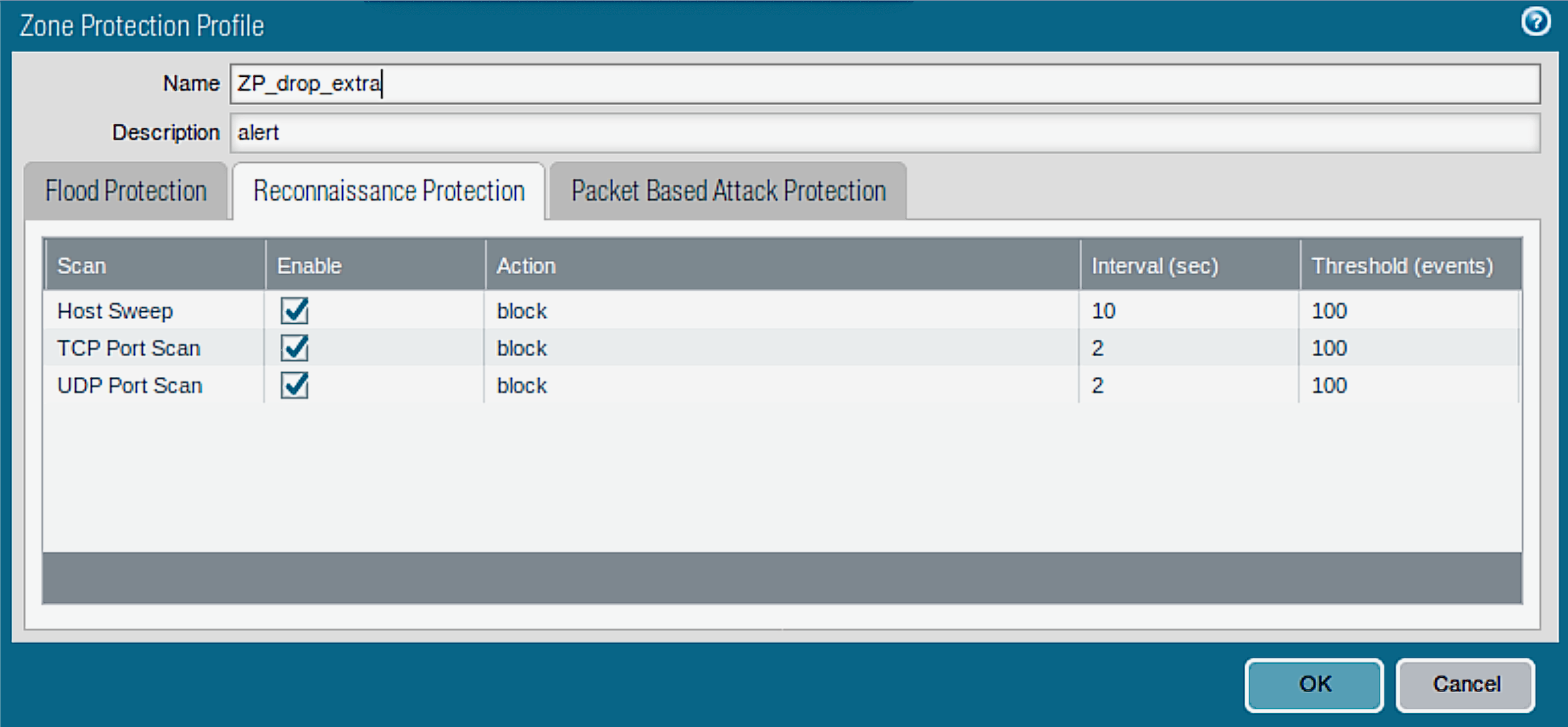}}\quad
				\label{fig:1D}
			}
		}
		\mbox{
			
			\hspace{-4mm}
			\subfigure[DDoS -- SYN flood.]{
				
				{\includegraphics[width=0.55\textwidth]{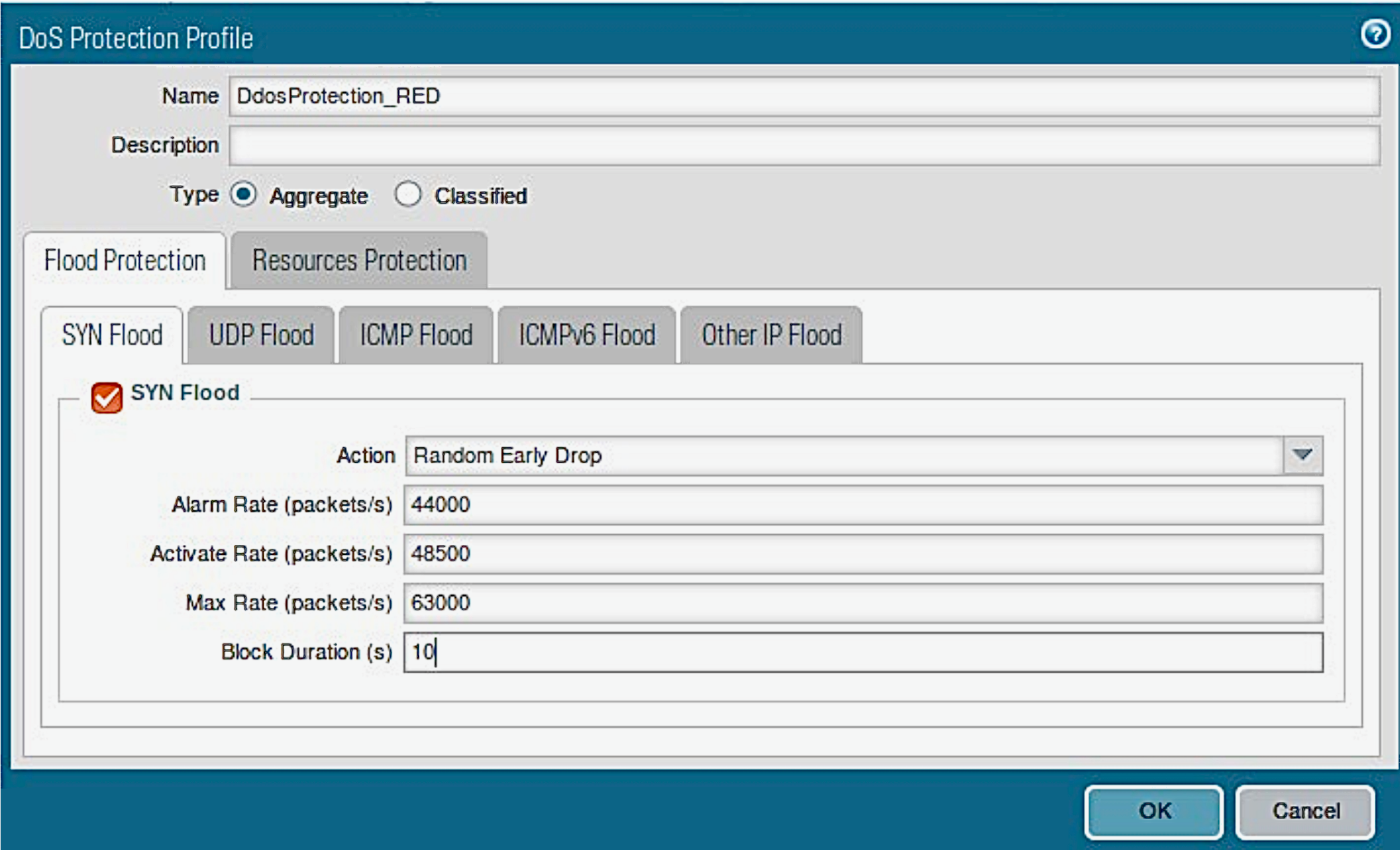}}\quad
				\label{fig:2D}
			}	
			\hspace{-4mm}
			\subfigure[DDoS -- UDP flood.]{
				{\includegraphics[width=0.41\textwidth]{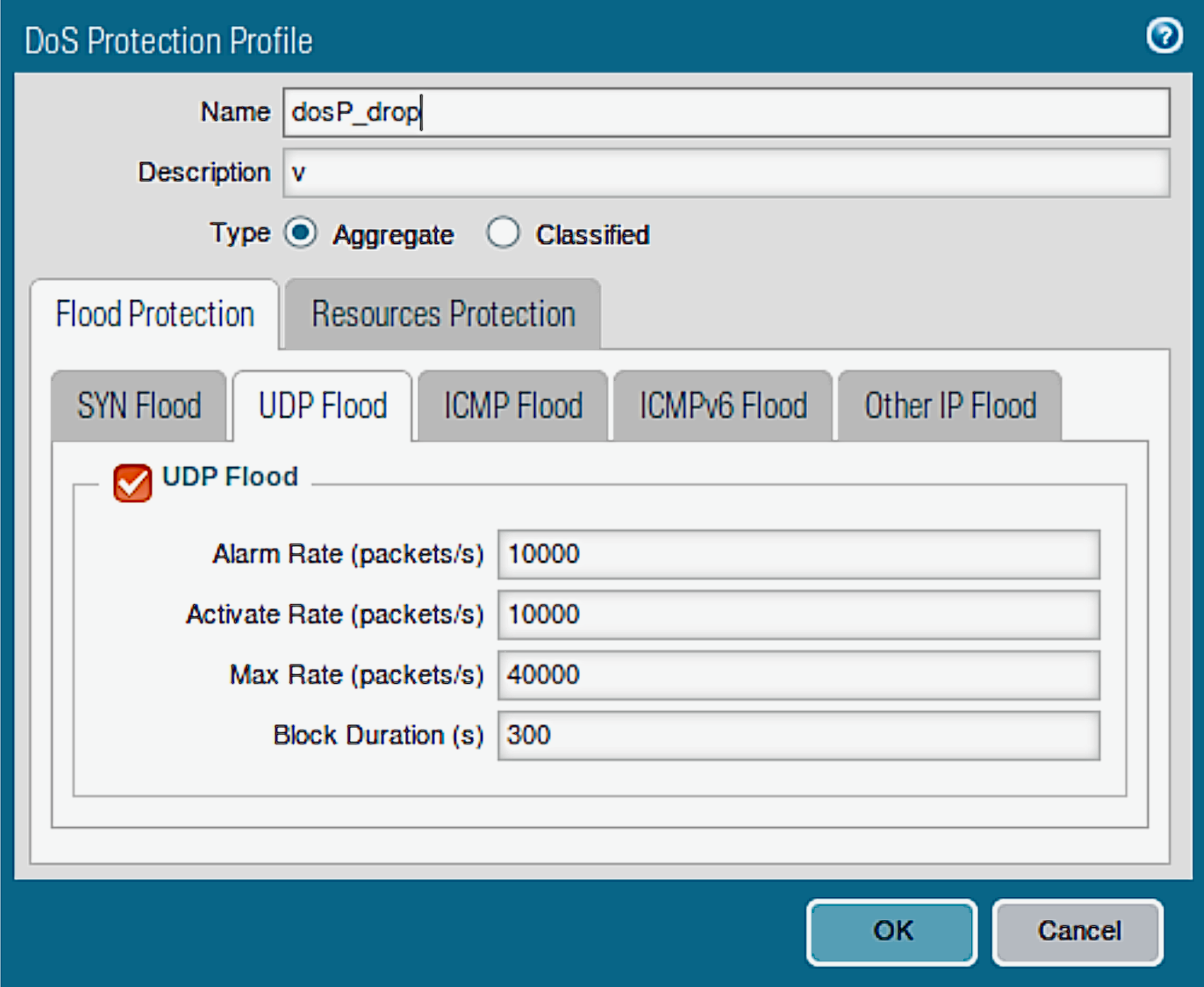}}\quad
				\label{fig:3D}
			}	
		}	
		
		\caption{Firewall configurations available for distributed network attack protection (\ie detection and mitigation): (a) reconnaissance/scan protection, (b) SYN flood DDoS protection, and (c) UDP flood DDoS protection.}
		\label{fig:3DScreenshot}
		\vspace{-3mm}
	\end{center}
\end{figure*}

\subsubsection{Thresholds in Commercial Appliances}
Proprietary appliances such as next-generation-firewall (NGFW), typically deployed at the border of enterprise networks, use threshold-based mechanisms for detecting attacks.
Network administrators configure rules to govern access policies of certain networked hosts. Each rule may specify thresholds on traffic volume (\eg packet rates to specified IP addresses) to distinguish normal and/or abnormal communications.
In Fig.~\ref{fig:3DScreenshot}, we show three screenshots of configuration pages on a commercial firewall appliance, often used in large-scale networks. It can be seen how defensive (default) rules against reconnaissance attacks (Fig.~\ref{fig:1D}), DDoS attacks via SYN flood (Fig.~\ref{fig:2D}), and DDoS attacks via UDP flood (Fig.~\ref{fig:3D}), are configured.
For the reconnaissance protection, shown in Fig.~\ref{fig:1D}, the network administrator who wants to protect their assets from host reconnaissance or port scans may set up a security rule to block all external IP addresses that send more than 100 packets to intended hosts within a specified interval (say, 2 or 10 seconds).
For protecting against DDoS via SYN flood in Fig.~\ref{fig:2D}, the administrator is able to configure thresholds on the packet rate of inbound TCP-SYN toward certain IP zones -- exceeding thresholds indicates volumetric anomalies, thereby triggering alerts or actions. Likewise, Fig.~\ref{fig:3D} shows similar detection and mitigation thresholds configured for UDP-based DDoS attacks.

Legacy proprietary middleboxes, enabling admin-configured rules, have been widely deployed by the industry for distributed attack detection/mitigation. Through, these methods are quite simple for adoption and relatively effective for certain attacks types, they are insufficiently flexible to fulfil emerging security needs like detecting distributed attack sources with versatile traffic patterns.
The absent standard way of configuring rules and policies across security appliances sourced from diverse manufacturers and a lack of full compatibility between commercial vendors will introduce practical challenges to operators of multi-vendor networks. It becomes difficult for them to effectively apply their detection logic across appliances, each protecting parts of their network \cite{KBordersSecurity2012}.

\subsubsection{Experimenting with Expressive Queries}
Given some problems discussed above, particularly a lack of flexibility in configuring rules, researchers \cite{SNarayanaSigcomm2017,PGaoSec2018,AGuptaSigcomm2018} employed programmable networking techniques to prototype an expressive query-based middlebox that allows for configuring reactive rules (essentially thresholds-based) at run-time.
For instance, to realize a sustainable and versatile attack detection mechanism, particularly in fast-changing environments, the authors of \textit{Marple} \cite{SNarayanaSigcomm2017} designed a query language to perform monitoring tasks via key-value store primitives on programmable P4 switches. 
To make a rule-based security mechanism effective in combating sophisticated cyber-attacks, involving various logical steps and targeting a large number of network entities, \textit{SAQL} \cite{PGaoSec2018} was developed as a stream-based query system that provides an anomaly query engine that allows users to specify their complex detection logic using domain-specific languages.
By leveraging both programmable P4 switches and software stream processors, \textit{Sonata} \cite{AGuptaSigcomm2018} was proposed as a network telemetry system that is scalable and expressive in performing security tasks (\eg detection of SSH brute force, port scan, DDoS, or Slowloris attacks) with fewer configurations compared to prior relevant systems.
Although those research ideas still have a long way to go before being fully adopted by the industry, they are valuable steps toward realizing a low-cost, easy-to-upgrade, and expressive rule-based detection system.

\subsubsection{Performance Evaluation and Rule Optimization}
While rule-based security systems are relatively prevalent across the computer networking industry, configuring effective and error-free specifications requires expert administrators with sufficient domain knowledge, as well as complete visibility into connected assets on their networks, without which they can hardly set up effective thresholds, queries, or take appropriate actions. 
Moreover, manually managing configurations can be challenging for medium to large enterprise networks with complex host compositions and behaviors, particularly in handling rule redundancies, logical conflicts, and configuration errors.

Optimizing the placements of security rules and identifying potential redundancies have received extensive attention from researchers.
The work in \cite{MRLyuCOMPSAC2000} conducted experiments (\eg with the number of rules and their complexity) to evaluate the performance degradation in latency and bandwidth that may be caused by placing firewall policies at various security levels.
The authors concluded that the placement of firewall rules can have significant impacts on metrics such as latency and throughput, thus, optimization of firewall technologies is critical in reducing performance losses.
The authors of \cite{AWoolComputer2004} conducted a quantitative analysis of rule sets and configuration errors available on a commercial firewall in production, highlighting that corporate firewalls are often improperly configured, which prevents them from providing sufficient security protection.
Although vendors supply templates and guidelines, network administrators often face challenges in manually selecting and efficiently adopting those templates for their networks.
To better understand the performance impacts of rule-based firewalls, the authors of \cite{KSalahTNSM2012} developed a queuing model with a Markov chain to model key performance metrics of firewalls when handling normal or DoS traffic flows. Work in \cite{CWangICIAS2014,MAimonJAMSI2015} extensively studied performance bottlenecks such as CPU and memory usage under network conditions such as varying traffic rates, packet sizes, and the number of communication flows.

With operational challenges and performance bottlenecks associated with rule-based solutions discussed above, various optimization methods have been employed.
Legacy firewalls check each received packet against individual existing rules. Therefore, increasing the number of firewall rules will unavoidably lead to larger processing time. 
The work in \cite{UMustafaIWCMC2013} proposed a data mining approach to predict hit probabilities of mutually exclusive rules so that they could be ordered based on their popularities, significantly reducing the processing time up to 40\%.
To tackle rule redundancies, the authors of \cite{AVasuIJCN2014} proposed an optimization algorithm to locate and reduce redundant rules configured on an enterprise firewall. Work in \cite{HTegenawAFRICON2015} designed a stateful firewall architecture that can classify network traffic according to their application types; each is mapped to a customized processing pipeline to achieve better performance in terms of CPU utilization, throughput, and queuing delay.
Work in \cite{PLeeTENCON2017} developed a hash-based packet classification algorithm to significantly reduce the delays caused by the rule-matching process on a typical firewall appliance. Although a handful of prior research works exist on optimally managing errors and performance degradation introduced by redundant firewall rules (manually configured), rule-based firewall performance issues are still key concerns yet to be solved \cite{FirewallPerformanceIssues2018}.

\subsection{Community Signature-Based Detection}\label{sec:signature-based}
With the increasing complexity of attack vectors, enforcing effective security rules by administrators has become more challenging than ever. To ease this pain point, the security community developed various software intrusion detection systems (\eg Bro \cite{VPaxsonCN1999} and Snort \cite{MRoeschLisa1999}) that do not require complex configurations. Instead, users could simply import security signature files containing fingerprints of malicious traffic characterized and made publicly available by security experts and/or researchers.

\subsubsection{Merits} 
Unlike rule-based detection via proprietary systems, signature-based attack detection typically leverages software engines (CPU-based computing) that support highly flexible traffic processing functions.
In addition, software-based intrusion detection systems (IDS) are relatively attractive because various functionalities can be customized by network/security admins without tedious negotiations with vendors to upgrade hardware appliances.
As highlighted in \cite{MZhangNDSS2020}, hardware appliances are designed for high performance (\eg sustained Tbps traffic) and thus sacrifice operational flexibility in dynamic network environments. At the same time, software-based systems can overcome those limitations by elastically scaling or replacing detection functions based on operational needs and traffic composition.

\subsubsection{Current Issues} 
Despite certain advantages software signature-based IDSes offer, there are practical challenges in using these tools in operation. 
First, generating quality signatures for diverse attacks can be nontrivial and time-consuming, requiring expert (and expensive) man powers. Second, given the attack surface of various networks can be quite different, signatures developed by third parties may not be readily (and directly) applicable to every enterprise network -- yielding poor efficiency. Third, such software tools hardly scale cost-effectively to process high traffic rates.
To address three problems, researchers developed methods for the automatic generation of signatures, increasing the efficacy of detection, and improving the scalability of software IDSes.

\textbf{Automatic Generation of Signatures:}
Many research works attempted to develop automatic methods for generating reliable attack signatures. Work in \cite{CKreibichCCR2004} presented a system to automatically generate signatures needed for pattern matching and protocol conformance checks. The authors set up honeypots to passively capture malicious network traffic. To evade getting matched against known signatures, attackers may try to craft the payload contents of their malicious packets. To defend against those sophisticated attacks, \textit{Polygraph} \cite{JNewsomeSP2005} was proposed to automatically generate signatures that contain multiple disjoint content sub-strings for polymorphic worms (\ie an example attack that varies its packet payloads frequently). \textit{AutoRE} \cite{YXieCCR2008} focuses on detecting those botnets that send spam emails. The authors aimed at avoiding allowlists which can be tedious to populate. They instead check whether email payloads contain identifiable malicious patterns URLs and look for distributed destination and/or bursty patterns in the email traffic sent.

\textbf{Detection Effectiveness:}
In terms of the prediction power of signature-based detection systems, researchers have identified various problems and proposed corresponding solutions. In \cite{SPattonRAID2001}, \textit{S. Patton et al.} highlighted the ``Squealing'' vulnerability of a signature-based IDS. Given known signatures employed by the IDS in charge, attackers can craft malicious packets that result in high rates of false positives making the alerting system almost unreliable (useless).	Authors of \cite{RSommerCCS2003} observed that the legacy signatures using byte sequences suffered from a high false-positive rate due to the dynamics of attacks. To address this issue, the authors developed a signature engine on the Bro IDS \cite{VPaxsonCN1999} that can generate richer signatures by incorporating factors like the dependency of networking events (\eg requests and replies). Works in \cite{DDayICDS2011} and \cite{EAlbinWAINA2012} compared the accuracy and performance of IDS designed for computing environments: single-threaded tools (\ie Snort) versus multi-threaded tools (\ie Suricata). They concluded that Suricata gives higher accuracy under a multi-core setup, while Snort achieved fewer false negative alarms within a single-core networking system.
Besides, according to \cite{NChaabouniCST2019,EBenkhelifaCST2018}, the adoption of emerging assets such as IoT and sensors makes legacy security signatures less effective in flagging malicious activities, as they exhibit different traffic patterns compared with typical IT networked hosts and assets. 

\textbf{Scalability:}
Software-based IDSes incur high computational costs and often do not scale well (unlike specialized hardware appliances) to handle high traffic rates cost-effectively \cite{MZhangNDSS2020}. Ineffective design of software components can make this problem even worse \cite{GLiuKBNets2017}.
Therefore, signature-based IDSes running software platforms are mostly used by relatively smaller enterprises with low traffic rates. Researchers incorporated various techniques to improve the scalability of software IDSes.
First, many prior works have exploited the concept of distributed computing.  \textit{The NIDS cluster} described in \cite{MVallentinRAID2007} used distributed computational nodes with optimized coordination approaches to achieve decent performance with software-based stateful intrusion detection.
The authors of \cite{DCarliCCS2014} proposed a domain-specific model that distributes traffic analysis across different processing units with specific functions to achieve scalability and efficient detection on multi-core hardware.
Also, there exist works that developed methods to reduce the overheads by signature-matching.
For example, work in \cite{SKongSecureComm2008} developed an alphabet compression table that combines distinct input signature symbols with identical behavior into one symbol, thereby reducing memory usage.
\textit{O3FA} \cite{XYuANCS2016} was proposed to achieve packet ordering and flow reassembly during pattern-matching phases with low buffer consumption, which is particularly useful in reducing computational overheads when handling attack traffic with long sequences of out-of-order packets.
Moreover, with the increasing popularity of virtualization technologies, network intrusion detection on virtualized platforms is proven to be useful in reducing overheads, as it supports dynamic scaling of computational resources and flexible deployment of detection functions. For example, in \cite{JDengNDSS2017}, \textit{J. Deng et al.} built a virtualized IDS regulated by a virtualized controller for semantic consistency, correct flow update, buffer overflow avoidance, and optimal scaling in real time. \textit{vNIDS} \cite{HLiCCS2018} employed a detection state-sharing mechanism to reduce the virtualization overhead of IDS. Therefore, it achieves elasticity in detecting attacks of various profiles and also guarantees acceptable scalability.

\subsection{Fine-Grained Detection using Flow Statistics}
Distributed attacks sourced from external botnets are often mixed with benign traffic flows from legitimate sources to the victim enterprise server \cite{MLyuCN2023}. However, both proprietary rule-based and community signature-based detection systems (discussed in \S\ref{sec:rule-based} and \S\ref{sec:signature-based}) barely maintain fine-grained traffic statistics for individual flows. Instead, they focus more on aggregate statistics (destination IP/subnet-level). Therefore, they face challenges in providing the necessary visibility for precisely differentiating malicious flows from benign ones destined for the victim, particularly when attack sources are distributed.
Many researchers have proposed methods for anomaly detection in network traffic using flow-level statistics, which enables them to achieve precise attack detection/mitigation without causing collateral damage \cite{CDietzelCoNEXT2018}, \ie dropping only packets in malicious flows without affecting packets in benign flows.

\subsubsection{Scalability Issues}
It is important to note that collecting and analysing fine-grained flow statistics across a large and fairly active network may not always be practical. Therefore, many research efforts have been made to develop lightweight data structures to maintain flow statistics.
\textit{Kronecker graph} \cite{JLeskovecJMLR2010} was designed to model network flows using graphs generated by a non-standard matrix operation called Kronecker product, which is both descriptive and practical.
The authors of \cite{JVilalobosBDCAT2017} leveraged distributed computing nodes that collectively maintain in-memory graphs containing flow statistics to detect DDoS attacks cost-effectively.
Many prior works employed streaming (online) algorithms to realize attack detection with relatively lower computational costs. 
\textit{STONE} \cite{MCallau-ZoriSAC2013} maintains traffic attributes pertaining to the volume of activities (\eg TCP SYN counts) for target asset groups. The authors employed streaming techniques that can scale and are more conducive to real-time monitoring.
Work in \cite{AMcGregorSIGMOD2014} systematically reviewed the processing methods, such as ``insert-only graphs'',  ``graph sketches'', and ``sliding window'', for streaming graphs that help to reduce the computational costs when processing flow statistics.
Work in \cite{DIppolitiTAAS2016} developed an anomaly detection scheme, looking for malicious flows such as DDoS and reconnaissance attacks. The proposed scheme aggregates flow alerts based on their similarities/correlations in five-tuple metadata to address the scalability.

\subsubsection{Identifying Important Features}
Identifying key predictive features from flow statistics for attack detection is another popular direction.
Principal Component Analysis (PCA) is a method, widely used, to determine which features are more influential in classification tasks. 
To exclude (or reduce the impact of) redundant and less relevant attributes, the authors of \cite{FIglesiasML2015} proposed a multi-stage feature selection method. They utilized lightweight filters and heavy regression models to extensively examine the importance of features in a progressive manner. Commonly used features for network anomaly detection were examined, and less than 40\% of them were found to be effective in attack detection.
The work in \cite{CMaAccess2019}  introduced five groups of descriptive features (\eg flow metadata features, sequence packet features, and general statistical features) of network flows. The authors demonstrated the efficacy of those attributes in detecting seven types of network attacks, including SSH patator, DDoS, and port scan.

\subsubsection{Statistical Learning Methods}
Developing statistical learning methods using flow characteristics for better attack detection has been explored by researchers.
For example, S. Jin \textit{et al.} discussed their work in detecting SYN flooding attacks using a covariance analysis model in \cite{SJinICC2004}. They showed that the model could effectively distinguish benign flows and malicious flows by profiling their TCP headers.
K. Lee \textit{et al.}	\cite{KLeeESA2008} applied clustering algorithms to a set of traffic features (\eg randomness of source and destination IP addresses) 
to differentiate DDoS traffic from normal communications.
The authors of \cite{HRahmaniCC2012} employ a statistical metric called ``total variance distance'' that quantifies  the similarity between flows, achieving better performance in detecting attack traffic than legacy methods.

\subsection{Highlights}

In this section, we categorized attack detection methods into three types including proprietary rule-based, community signature-based, and fine-grained flow statistic-based detection.

Currently, the industry (at least large enterprises) widely adopts proprietary rule-based detection appliances for their ease of deployment and scalable operation. However, such an approach becomes less effective in combating dynamic attack vectors applied to expanded attack surfaces. It falls short of expectation, particularly at scale, when the enterprise network serves diverse asset classes and functionalities.

Signature-based detection is often realized as software products, relying on knowledge (\ie signatures) supplied by open-source communities. Optimally selecting appropriate signatures, developed by security experts, could help (to a great extent) network operators (of medium/smaller enterprises) to quickly respond to emerging cyber threats, (\eg ``Log4Shell exploit'' \cite{EDouglasSDC2022,SID58813,ELeblond2021}). However, appropriately setting up the software environments, routinely updating signatures, and, importantly, trusting the open-source community may not always be feasible for administrators of large organizations. Besides, they are often packaged as software tools on commodity servers that make them expensive to scale for a large network.

Network attack detection methods, leveraging fine-grained flow statistics, have proven their superiority in precisely identifying victims, attackers, and malicious flows of a distributed attack. However, real-time maintenance and processing of fine-grained traffic statistics for many concurrent flows traversing an enterprise network can hardly scale. Therefore, achieving scalability while not compromising the quality of visibility for flow statistic-based methods is a crucial challenge to address before they can be widely adopted.
\section{Opportunities of Emerging Paradigms \\for Network Security}\label{sec:SurveyEmergingTechniques}
The advancements in programmable networking and machine learning (ML) techniques have opened up new possibilities for addressing current challenges in enterprise network security. Researchers have recently leveraged these two specific technologies in various network security domains. For instance, they have developed orchestration systems that offer flexible attack detection capabilities in ISP networks (\S\ref{sec:programmableNetwork}) and proposed accurate algorithms specifically designed to detect certain types of attacks (\S\ref{sec:machineLearning}).

These seminal prior works serve the research community with foundational lessons in developing practical and effective security systems for large enterprise networks. In the subsequent sections, we will delve into the existing research in the areas of network security that utilize programmable networks and ML techniques, respectively.

\subsection{Programmable Networking for Network Security}\label{sec:programmableNetwork}
The concept of programmable networking, broadly speaking, stems from technologies like Network Function Virtualization (NFV) \cite{LNobachCCR2016} and Software-Defined Networking (SDN) \cite{NFeamsterTheRoadtoSDN2013} and enables flexible network monitoring and controls. These technologies empower IT and cyber teams to dynamically configure and update flow rules and/or network functions, allowing for custom security measures and defense utilities in response to their ever-changing attack surfaces.

\subsubsection{Practical Challenges}
While programmable networking sounds promising in enhancing defense capabilities, its adoption faces several practical challenges \cite{SSezerCM2013}, including the performance bottleneck of software controllers, specific vulnerabilities associated with controllers ad switches, scalability limits of software-based network functions, and concerns about compatibility with existing systems and middleboxes.

To address these challenges, researchers have made efforts to develop practical solutions.
For example, R. Sommer \textit{et al.} \cite{RSommerCCPE2009} proposed a specialized NFV architecture that effectively utilizes multi-core processors to achieve scalable network intrusion detection.
\textit{O3FA} \cite{XYuANCS2016} is developed as a lightweight packet inspection engine using deterministic finite automaton (\ie a finite-state machine) that processes out-of-order packet streams without reassembling flows. Thus, the system requires less memory than other packet inspection engines.
\textit{StateAlyzr} \cite{JKhalidNSDI2016} identifies and reduces the unnecessary operational processes for state clones in security middleboxes to achieve low computational overheads.
\textit{NetBricks} \cite{APandaOSDI2016} employs a zero-copy software isolation mechanism that significantly reduces the computational overheads in CPU and RAM usage on typical NFV platforms.
The NFV framework \textit{OpenNetVM} \cite{WZhangHotMIddlebox2016} is designed with high-level abstractions, allowing users to quickly build and deploy customized network functions without the need to handle complex optimization of computing resource allocations.
The hybrid packet processing pipeline \textit{ParaBox} \cite{YZhangSOSR2017} is specifically designed to incorporate parallel network functions, resulting in superior performance compared to traditional serial function chaining mechanisms.
\textit{StatelessNF} \cite{MKablanNSDI2017} breaks down virtual network functions into two components: a state management component to store stateful traffic information and a stateless packet processing component to extract packet information at high speeds. They are well separated and orchestrated by SDN utilities so that the traffic is processed in a more scalable manner.
\textit{vNIDS} \cite{HLiCCS2018} tackles the challenges of inefficient (and costly) detection of SDN/VNF-based systems by developing techniques such as state sharing among detection modules and dynamic slicing of detection logic programs.

\subsubsection{Prototypes for Attack Detection}
In addition to the research efforts to develop practical methods, prototypes have been built for certain attack detection problems that utilize programmable networks. These prototypes showcase the potential of programmable networking in enhancing security measures.

For example, R. Braga \textit{et al.} \cite{RBargaLCN2010} developed a system that utilizes programmable switches to extract flow statistics for detecting flooding attacks.
S. Lim \textit{et al.} \cite{SLimICUFN2014} utilized OpenFlow-based switches to achieve flexible isolation of bots in DDoS attacks.
In a study by K. Giotis \textit{et al.} \cite{KGiotisCN2014}, the authors developed a system that combines OpenFlow and sFlow utilities to collect and process network statistics for scalable anomaly detection.
The \textit{FlowTags} system \cite{SFayazbakhshNSDI2014} employs an SDN architecture to achieve flexible security enforcement through middleboxes at a network level with relatively low computational overheads.
The \textit{Bohatei} system \cite{SKFayazSec2015} utilizes SDN proactive and reactive flow rules to dynamically orchestrate network traffic forwarding through backbone networks, diverting attack traffic to be handled by specialized security middleboxes with appropriate computational resources.
C. Yoon \textit{et al.} \cite{CYoonCN2015} demonstrated the feasibility of utilizing programmable networks for cybersecurity by developing representative security functions of in-line firewalls, passive IDS, and network anomaly detectors with SDN technology.
The \textit{Atlantic} system \cite{ADasilvaNOMS2016} leverages the flexibility of SDN to detect, classify, and mitigate malicious flows in relatively small networks (\eg consisting of 100 hosts and two switches).
In \cite{CLiuSOSR2017}, the authors utilize SDN reactive routing to selectively forward only the initial packets of each network flow for deep packet inspection.
J. Deng \textit{et al.} \cite{JDengNDSS2017} constructed a virtual firewall architecture using SDN and NFV, enabling elastic rule placement and flexible detection functionalities.
\textit{Sonata} \cite{AGuptaSigcomm2018} achieves scalable traffic processing by offloading resource-intensive and repetitive network functions from software processors to hardware programmable switches.
\textit{ACC-Turbo} \cite{AAGranSigcomm2022} implements an in-network mechanism on a programmable P4 switch to detect DDoS attacks with short and high-rate pulse patterns.
Lastly, \textit{PEDDA} \cite{MLyuCN2023} utilizes a programmable control-plane switch (\ie OpenFlow) and virtual network functions to dynamically apply DDoS detection modules, each with specific capabilities and costs, enabling fine-grained detection and scalable real-time operation.

\begin{table*}[!t]
	\caption{Prior relevant surveys with different objectives and focuses.}
	\centering
	\renewcommand{\arraystretch}{1.2}
	\begin{tabular}{|l|l|l|c|}
		\hline
		\rowcolor[rgb]{ .906,  .902,  .902}		\textbf{Survey} & \textbf{Key objectives}                                                 & \textbf{Focused aspects}                      & \textbf{The latest year of article reviewed} \\ \hline
		{\cite{JMirkovicCCR2004}}  & DDoS                                                             & Attack and defense mechanism                   & 2003                                \\ \hline
		{\cite{PGarciaTeodoroCS2009}}  & Anomaly-based Network Intrusion Detection                        & Attack and defense mechanism                & 2005                                \\ \hline
		{\cite{STZargarCST2013}} & DDoS                                                             & Attack and defense mechanism                   & 2012                                \\ \hline
		{\cite{JJangJCSS2014}}  & Typical Protection Methods for IT Infrastructure                 & Attack and defense mechanism & 2013                                \\ \hline
		{\cite{ABuczakCST2016}}  & Intrusion Detection via Data Mining and Machine Learning & Dataset and defense mechanism      & 2015                                \\ \hline
		{\cite{MUmerCS2017}} & Flow-based Intrusion Detection                                   & Dataset and defense mechanism      & 2016                                \\ \hline
		{\cite{AVoronkovCS2017}} & Network Firewall                                                 & Configuration method       & 2016                                \\ \hline
		{\cite{ATundisARES2018}} & Network Vulnerability Scanning                                   & Attack and defense mechanism               & 2017                                \\ \hline
		{\cite{KNeupaneSC2018}}  & Next-Generation-Firewall                                         & Defense mechanism    & 2015                                \\ \hline
		{\cite{CChenAIS2019}}  & Network Situational Awareness                                    & Defense mechanism                  & 2018                                \\ \hline
		{\cite{SSenguptaCST2020}} & Moving Target Defense                                            & Defense mechanism                  & 2019                                \\ \hline
		{Our survey} & Asset Monitoring and Distributed Attack Detection          & Attack and defense mechanism                & 2023                            \\ \hline
	\end{tabular}
	\label{tab:RelatedSurveyDefense}
\end{table*}

\begin{table}[t!]
	\caption{Relevant surveys on different network types.}
	\renewcommand{\arraystretch}{1.2}
	\begin{tabular}{|l|l|c|}
		\hline
		\rowcolor[rgb]{ .906,  .902,  .902}		\textbf{Survey} & \textbf{Network type}             & \textbf{The latest year of article} \\ \hline
		{\cite{XChenCST2009}}                                              & Wireless sensor network  & 2007                                \\ \hline
		{\cite{SScott-HaywardCST2016}}                                             & Software-defined network & 2014                                \\ \hline
		{\cite{IAhmadCST2015}}                                               & Software-defined network & 2015                                \\ \hline
		{\cite{NChaabouniCST2019}}                                              & IoT network              & 2018                                \\ \hline
		{\cite{EBenkhelifaCST2018}}                                              & IoT network              & 2017                                \\ \hline
		{\cite{NAgrawalCST2019}}                                               & Cloud networks  & 2018                                \\ \hline
		{\cite{JCaoCST2020}}                                              & 3GPP 5G network          & 2019                                \\ \hline
		{\cite{ABhardwajCSR2021}}           & Cloud networks & 2021 \\ \hline
		Our survey                                            & Enterprise network       & 2023                                \\ \hline
	\end{tabular}
	\label{tab:RelatedSurveyNetworkType}
\end{table}

\subsection{Machine Learning for Network Security}\label{sec:machineLearning}
Machine learning techniques have proven their efficacy in accurate inference (classification and/or anomaly detection) in domains like computer vision and speech recognition. Although the application of machine learning in cybersecurity faces practical challenges \cite{PGarciaTeodoroCS2009, RSommerSP2010}, researchers have made significant progress in developing machine learning-based methods and systems to enhance the security of various networks \cite{ABuczakCST2016}. These relevant prior works are wide in scope and objectives, providing valuable insights to the research community.

For example, the \textit{MADAM ID} framework, proposed in \cite{WLeeTISS2000}, utilizes machine learning-based data mining techniques to process network telemetry data (\eg packet and flow events and connection status) for intrusion detection.
In \cite{JShunICNC2008}, J. Shum \textit{et al.} employed simple neural networks trained with back propagation algorithms for detecting network attacks such as DDoS, spam, and exfiltration.
The \textit{BotMiner} system \cite{GGuSecurity2008} applied unsupervised clustering algorithms to characterize the behavior of botnet groups that exhibit similar patterns in their command-and-control activities. Such similarity can be determined in traffic attributes like the number of flows generated per hour.
M. Lyu \textit{et al.} utilized clustering algorithms in \cite{MLyuTNSM2022} to classify enterprise DNS assets and health metrics based on their DNS traffic profiles for anomaly detection.
In \cite{LKocESA2012}, L. Koc \textit{et al.} introduced the Hidden Naive Bayes (HNB) method for network intrusion detection, outperforming other machine learning models in handling high-dimensional data, identifying dependent features, and reducing computational overheads.
A scheme designed by M. Javed \textit{et al.} in \cite{MJavedCCS2013} specifically focuses on detecting SSH brute-forcing attacks using a beta-binomial distribution model.
The authors in \cite{AShrivasIJCA2014} developed an ensemble model that combines Bayesian Network with Gain Ratio for feature selection and Artificial Neural Network for attack detection. C. Hsieh \textit{et al.} proposed a DDoS detection system in \cite{CHsiehICASI2016} that employs neural networks on Apache Spark big data computing clusters to handle high data rates traffic. \textit{DeepLog} \cite{MDuCCS2017} employs deep learning algorithms for detecting anomalies in system logs collected from enterprise hosts. H. Siadati \textit{et al.} used machine learning-based algorithms to identify anomalous logins within an enterprise network \cite{HSiadatiCCS2017}. In \cite{DTangFGCS2022}, D. Tang \textit{et al.} developed data-driven models to detect relatively low-rate DoS attacks that exhibit abnormal patterns in the frequency, variation, and distribution of TCP flows.

It is worth noting that most existing works focus on developing high-accuracy models and algorithms to detect specific types of attacks. These efforts demonstrate the effectiveness of machine learning-based methods in addressing network security challenges. Furthermore, these advancements lay the foundation for developing data-driven solutions in asset management and distributed attack detection, offering tailored approaches to large enterprises.

\subsection{Highlights}
In this section, we have summarized the latest advancements in, as well as the adoption of programmable networking and machine learning techniques in the field of network security.

The programmable networking paradigm offers flexible and dynamic traffic forwarding and measurement capabilities, outperforming traditional systems. That said, there are still practical challenges to be addressed before a wide adoption is realized. Key challenges include scalability issues arising from limited resources in both control and data planes (\eg switch memory size and flow entries), the need for skilled administrators comfortable with coding, and compatibility with existing network infrastructures.
Existing methods empower network and cyber teams to adjust the visibility levels and the granularity of network telemetry for real-time security inferences. Known techniques allow for collecting precise, fine-grained statistics based on the network size, composition of connected assets, and the evolution of attacks.

Machine learning techniques (data-driven models) promise to automatically classify hosts' traffic or detect attacks by trained network data models instead of relying solely on manually defined thresholds and/or signatures. However, applying machine learning algorithms to network security requires the community to overcome certain challenges. These barriers include striking a balance between descriptive attributes and scalability, handling false positives that lead to operational implications, and ensuring the explainability of inferences. Careful consideration of these challenges is essential for successfully adopting machine learning-based solutions in network security.

\section{Discussion on Research Gaps}\label{sec:discussion}
After extensively reviewing current techniques of asset behavior monitoring and distributed attack detection for enterprise networks, we found several open issues that require further investigation in future work.

\subsection{Dynamic and Scalable Host Monitoring} 
The complexity of networked assets and the dynamic nature of their communication patterns pose challenges for legacy methods that rely on static configurations and inferences. While dynamic networked graphs can effectively capture host behavioral profiles, maintaining such graphs for a large and high data- rates network becomes infeasible. Therefore, there is a need to develop methods that can achieve scalable monitoring of assets while ensuring dynamic and fine-grained visibility into necessary traffic portions. This represents a valuable future direction in network security research.

\subsection{Role-Aware Network Attack Detection} 
Every networked asset can be a victim of distributed attacks. We note that different assets come with relatively distinct communication patterns and vulnerabilities. Therefore, an effective defense system demands some form of customization in the monitoring techniques based on types of assets continence to the network.
Presently, attack detection appliances often apply generic detection mechanisms to the entire network or, at best, rely on some manual configurations by the network administrators, who manage policies for specific hosts or network segments. However, this approach may not fully incorporate the distinct characteristics and vulnerabilities of individual enterprise hosts. Hence, a valuable contribution would involve the development of automatic configurations for attack detection mechanisms utilizing the traffic profiles of enterprise hosts. By analyzing these profiles, automated configuration methods can dynamically tailor detection mechanisms to align with each host's unique requirements and attributes.

\subsection{Explainable ML-Based Attack Detection} 
Despite the promising prediction quality offered by machine learning (ML) methods for tasks like host classification and attack detection in controlled environments, their performance in operational networks can be unknowingly impacted by various factors, such as limited training data, imperfect statistical features, and algorithmic biases, or concept drifts. In order to use the predictions from (black-box) ML models for high-stakes decision-making, network operators may require some assurance, assistance, or at least an explanation that helps them interpret and analyze specific inferences made by trained models.   
This is particularly important to avoid mishandling false-positives, leading to unnecessary disruptions and resource wastage.

\subsection{Self-Driving Enterprise Security Systems}

The complexity of configuring current network security and management systems poses challenges to the IT and cyber departments of large enterprises. These systems often rely on manual configurations and tweaks to specify intents like which assets/segments receive priority for protection, protecting against which types of threats (\eg scans or DDoS), setting detection thresholds, and choosing appropriate mitigation actions. Managing numerous and complex policies can be cumbersome and prone to human errors. Additionally, sub-optimal configurations (\eg inconsistent and conflicting rules) can compromise the security and stability of networks. To prevent or at least manage these risks and improve operational efficiency, there is a need to explore developing ``self-driving'' security systems. These systems are expected to operate (to a great extent) automatically, gradually becoming autonomous and independent of manual configurations.

\section{Related Surveys on Network Security}\label{sec:RelatedWorks}
We now discuss some of the existing survey papers that focused on different aspects of network security.

\subsection{Attack Detection Methods} 
A group of literature reviews focuses on categorizing methods for detecting network attacks. Their key objectives, focused aspects, and the latest year of articles reviewed by those surveys are summarized in Table~\ref{tab:RelatedSurveyDefense}. The features of our study are captured in the last row of this table.

J. Mirkovic \textit{et al.} \cite{JMirkovicCCR2004} provided a taxonomy of DDoS attacks and  corresponding defense mechanisms.
The authors of \cite{PGarciaTeodoroCS2009} categorized the system architecture of underlying modules inside network intrusion detection systems (NIDS).
S. T. Zargar \textit{et al.} \cite{STZargarCST2013} highlighted DDoS defense mechanisms with a focus on where on the network they are applied and when defense actions take place.
The authors in \cite{JJangJCSS2014} comprehensively discussed vulnerabilities in the networking ecosystem targeted by emerging cyber-attacks and their countermeasures.
The types and mechanisms of data mining and machine learning methods and their applications in cyber-security research have been discussed in \cite{ABuczakCST2016}.
Work in \cite{MUmerCS2017} summarizes flow-based intrusion detection techniques, datasets, and prototypes.
A. Voronkov \textit{et al.} \cite{AVoronkovCS2017} thoroughly reviewed the usability aspect of firewall configurations.
A. Tundis \textit{et al.} \cite{ATundisARES2018} reviewed existing vulnerability scanner tools applied for benign or malicious purposes.
The authors in \cite{KNeupaneSC2018} discussed the functionalities of popular next-generation firewalls (NGFW) and their efficacy in coping with emerging network threats.
C. Chen \textit{et al.} \cite{CChenAIS2019} focused on the architecture of situational awareness systems for network security, which include data collection, situational understanding, prediction, and visualization.
S. Sengupta \textit{et al.} \cite{SSenguptaCST2020} comprehensively discussed the effective methods to defend against attacks originating from moving targets.

\subsection{Specific Attacks on Certain Network Types}

A cluster of survey papers studied attacks specific to certain network types, which are more vulnerable given their distinct characteristics. Table~\ref{tab:RelatedSurveyNetworkType} summarizes these survey papers, showing their focus and the latest year of articles reviewed. The features of our survey are shown in the last row of this table.

X. Chen \textit{et al.} \cite{XChenCST2009} summarized security problems in wireless sensor networks and discussed the efficacy of existing defense techniques.
The authors in \cite{SScott-HaywardCST2016} and \cite{IAhmadCST2015} highlighted security issues of software-defined networks and provided a list of key requirements for an effective defense architecture.
Works in \cite{NChaabouniCST2019} and \cite{EBenkhelifaCST2018} focused on network intrusion detection for IoT networks.
N. Agrawal \textit{et al.} \cite{NAgrawalCST2019} particularly focused on defense mechanisms against DDoS attacks for cloud computing networks.
J. Cao \textit{et al.} \cite{JCaoCST2020} summarized the security challenges, requirements, and gaps in 3GPP 5G networks.
A. Bhardwaj \textit{et al.} \cite{ABhardwajCSR2021} surveyed solutions developed by academia and industry to combat DDoS attacks on cloud networks.

\subsection{The Focus of our Survey} 
Prior surveys primarily categorized certain attack types (\eg DDoS) and corresponding defense methods depending on aspects such as target locations, attacking techniques, and the exploited network vulnerabilities. In contrast, our survey focuses on studying a broad range of distributed network attacks (not limited to DDoS),  countermeasure techniques, and opportunities promised by emerging paradigms, specifically for enterprise networks. 
Our survey reviews relevant research papers published until 2023 and provides valuable insights into unique challenges and opportunities for enterprise network security. This survey fills a gap in the existing literature by providing a comprehensive reference specifically tailored to the needs of researchers and practitioners working in enterprise network security.

\section{Conclusion}\label{sec:conclusion}
This survey focused on distributed network attacks on enterprise-connected assets and highlighted various countermeasures, including asset monitoring and attack detection systems. We discussed two types of distributed attacks (reconnaissance and DDoS) on enterprise assets. We reviewed existing methods (developed by academia and industry) for monitoring the behaviors of enterprise hosts and detecting distributed attacks. We highlighted the capabilities of two emerging/rising technologies (\ie programmable networks and ML) that bring new opportunities in addressing enterprise network security concerns. Lastly, we highlight several open issues as valuable future directions that are worthwhile to be explored. This paper provides a solid reference and inspires future research addressing enterprise network security issues.

\bibliographystyle{IEEEtranS}
\bibliography{IEEEabrv,ReferencesSurvey}

\end{document}